\documentclass[journal]{IEEEtran}

\usepackage{amssymb}
\usepackage{amsmath}
\usepackage{graphicx}
\usepackage{setspace}
\usepackage{color}
\usepackage{lipsum}
\usepackage{balance}
\usepackage{flushend}

\begin{document}
%
\title{Application of Expurgated PPM to Indoor Visible Light Communications - Part I: Single-User Systems}

\author{{Mohammad Noshad,~\IEEEmembership{Student Member,~IEEE}, and Ma\"{\i}t\'{e} Brandt-Pearce,~\IEEEmembership{Senior Member,~IEEE}}
\thanks{Mohammad Noshad (mn2ne@virginia.edu) and Ma\"{\i}t\'{e} Brandt-Pearce (mb-p@virginia.edu) are with Charles L. Brown Department of Electrical and Computer Engineering, University of Virginia Charlottesville, VA 22904.}
}

\markboth{}{Shell \MakeLowercase{\textit{et al.}}: Bare Demo of IEEEtran.cls for Journals}

\maketitle

\begin{abstract}
Visible light communications (VLC) in indoor environments suffer from the limited bandwidth of LEDs as well as from the inter-symbol interference (ISI) imposed by multipath. In this work, transmission schemes to improve the performance of indoor optical wireless communication (OWC) systems are introduced.
Expurgated pulse-position modulation (EPPM) is proposed for this application since it can provide a wide range of peak to average power ratios (PAPR) needed for dimming of the indoor illumination. A correlation decoder used at the receiver is shown to be optimal for indoor VLC systems, which are shot noise and background-light limited. Interleaving applied on EPPM in order to decrease the ISI effect in dispersive VLC channels can significantly decrease the error probability. The proposed interleaving technique makes EPPM a better modulation option compared to PPM for VLC systems or any other dispersive OWC system. An overlapped EPPM pulse technique is proposed to increase the transmission rate when bandwidth-limited white LEDs are used as sources.

\end{abstract}

\begin{IEEEkeywords}
Indoor optical wireless communications, visible light communications, optical networks, expurgated pulse position modulation (EPPM).
\end{IEEEkeywords}

\section{Introduction}\label{sec:intro}
\IEEEPARstart{O}{PTICAL} wireless communications (OWC) has been recently proposed for indoor networks in order to provide high-speed access for phones, PCs, printers and digital cameras in offices, shopping centers, warehouses and airplanes \cite{Indoor-OWC-Haas11}. The integrability of visible light communications (VLC) with the illumination system and its interference-free nature are the main factors that make this technique appealing for indoor applications. These systems must not only fulfill the requirements of the communication systems, but should also be able to support the indoor lighting needs. Yet the low-bandwidth optical sources used and dispersive channels encountered limit the throughput. This paper is Part I of a two-part paper exploring transmission techniques using expurgated pulse-position-modulation (EPPM) for high-rate indoor VLC systems \cite{EPPM12}. Part I addresses single-user communications while Part II proposes multiple-access techniques for multiuser networks. 

Dimming is an important feature of indoor lighting systems through which the illumination level can be controlled. A practical VLC system should support various optical peak to average power ratios (PAPR) so that for a fixed peak power the average power, which is proportional to the illumination, can be regulated. Continuous current reduction (CCR) and pulse-width modulation (PWM) are two techniques that have been proposed for dimming in indoor VLC systems \cite{OWC-Dimming-OFDM12}; these techniques require large bandwidths, and are therefore not suitable for high-rate systems.

In VLC systems, white light emitting diodes (LED) are used as sources, and modulation schemes that can be used in these systems are limited. Optical frequency division multiplexing (OFDM) and pulse position modulation (PPM) have been proposed for indoor wireless communications \cite{Indoor-OWC-Haas11}. OFDM is a spectrally efficient technique able to provide high-speed data transmission in dispersive communication systems. The application of OFDM to VLC systems with white LEDs is presented in \cite{OWC-OFDM-Haas09}. For a practical systems, the nonlinearity of the LEDs limits the performance of OFDM \cite{Indoor-OWC-Haas11}, and,  moreover, the possibility of controlling the PAPR in OFDM systems remains unanswered.

PPM is another option proposed for indoor OWC systems \cite{OWC-Coded-PPM12}. PPM is an appealing modulation scheme for optical communications since it requires simple transmitter/receiver structures and is easy to implement. However, it introduces several weaknesses in high-rate free-space optical systems requiring higher-order modulation, such as a high PAPR and low spectral efficiency \cite{EPPM12}. To alleviate these weaknesses, generalized forms of PPM have been proposed to improve its performance in optical systems. Overlapped-PPM (OPPM) is used to increase the data-rate of PPM in communication systems with bandwidth limited sources \cite{OPPM-84}. The application of OPPM to VLC systems is discussed in \cite{OWC-OPPM-PWM12}, where it is also used in combination with PWM to satisfy the average power constraint. Variable PPM (VPPM) is another modified form of PPM that is proposed in the IEEE 802.15.7 standard for visible light communication. It uses binary PPM to send data and changes the pulse-width to control the dimming level. While VPPM can provide flexible dimming compared to PPM, it has a lower spectral-efficiency compared to PPM for the same PAPR. Moreover, the overlapping technique cannot be used on VPPM to increase the data-rate for band-limited sources.

A generalized form of PPM called expurgated PPM (EPPM) is introduced in \cite{EPPM12} to decrease the bit-error probability in optical systems with power-limited sources. A multilevel form of EPPM is also proposed to increase the spectral-efficiency of PPM and improve its performance in bandwidth limited channels \cite{Multilevel-EPPM12}.
In this paper we propose to apply EPPM and multilevel-EPPM (MEPPM) to indoor VLC systems. These two modulations enable the system to operate over a wide range of PAPRs by controlling the ratio of the code-weight to the code-length of the generating code.

We introduce new approaches to make EPPM suitable for indoor-optical systems. Interleaving is applied on EPPM and MEPPM to reduce the interference between the adjacent time-slots in dispersive VLC channels and increase the data-rate. A lower bound is presented to show the performance limit of the best interleaver. The performance of the interleaved EPPM is evaluated for VLC systems that have a line-of-sight (LOS) reception and for those that receive only non-line-of-sight (NLOS) signals due to shadowing.
Overlapping the pulses in EPPM and MEPPM can provide high transmission-rates for white LEDs that have a limited bandwidth. The performance of overlapped and interleaved EPPM and MEPPM is studied for single-user VLC systems. The application of these two techniques can be generalized to any other peak-power limited communication system with restricted channel and/or source bandwidth.

The rest of the paper is organized as follows. In Section~\ref{sec:system}, we describe the indoor VLC systems. In Section~\ref{sec:approaches}, two techniques are proposed to increase the data-rate of optical communication systems, code interleaving and pulse overlapping. A lower bound on the performance of interleaved EPPM is calculated, and simulation results are presented to analyze the bit-error rate (BER) of the two approaches. Section~\ref{sec:conclusion} concludes the paper.

\section{System Description}\label{sec:system}

    \begin{figure} [!t]
    \begin{center}
    {\includegraphics[width=3.3in]{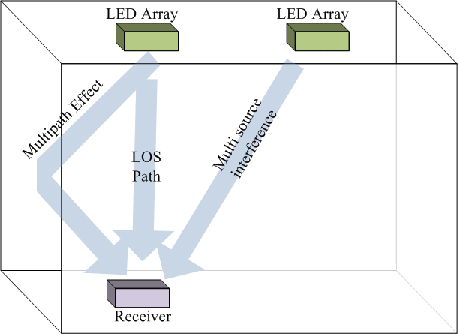}}
    \end{center}
    \vspace*{-0.1 in}
    \caption{The configuration of a VLC system with two lighting sources and one receiver.}
    \label{VLC}
    \vspace*{-0.1 in}
    \end{figure}

We focus on indoor optical systems comprising an LED array optical source, a diffuse channel, and a p-i-n photodetector, as described in this section.
LEDs are the preferred light sources for indoor communication applications due to  eye-safety regulations, low cost, and high reliability compared to lasers, incandescent and fluorescent sources \cite{Indoor-OWC-Haas11}. Using LEDs, both lighting and communications needs can be fulfilled at the same time. In a VLC system, where arrays of white LEDs are used as sources, the downlink configuration is as shown in Fig.~\ref{VLC}.  One or several sources are placed on the ceiling, and receivers are situated within the room, usually on desks. In our system, VLC is used only for the downlink, and for the uplink channel another system such as infrared (IR) communications can be used. In this way full-duplex communication does not cause self-interference.  The standard office illuminance determined by the Illuminating Engineering Society of North America at a distance of 0.8 m from the ceiling is 400 lx \cite{Illuminance-Standards}. For a detector with 1 cm$^2$ effective area, 1 lx generates $10^{-4}$ mW optical power.

Two technologies are commonly used to implement white light LEDs used for lighting applications. In the first technique, blue light is generated inside the LED, and then a layer of yellow phosphor is used to absorb the blue light and emit white light. The second type of white LEDs is a trichromatic LED, in which green, blue and red lights are generated independently inside the LED, and then combined to get white light. This kind of LED enables easy color rendering by adjusting each color independently. Despite the phosphorescent LEDs having a lower price compared to trichromatic LEDs, the latter are preferred for communication applications since they have a faster rise-time. In this work we utilize the latter.
%
%
While the bandwidth of the phosphorescent LEDs is limited to few MHz, that of the trichromatic LEDs can be as high as 20 MHz per color \cite{Indoor-OWC-Haas11}.

\begin{figure} [!t]
    \begin{center}
    {\includegraphics[width=3.3in]{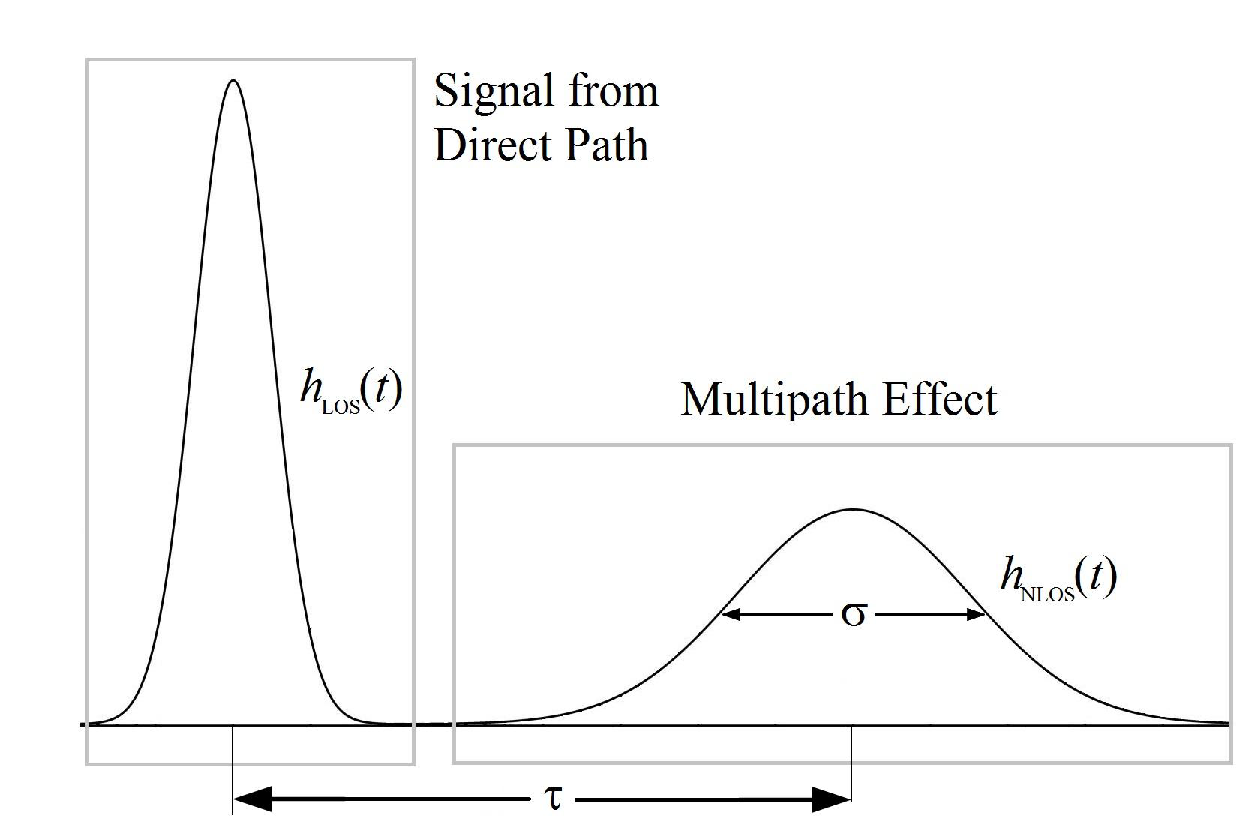}}
    \end{center}
    \vspace*{-0.1 in}
    \caption{The impulse-response of a VLC channel composed of  LOS and NLOS parts.}
    \label{VLC-Response}
    \vspace*{-0.15 in}
    \end{figure}

According to the results of \cite{OWC-Channel05} and \cite{OWC-Channel11} the impulse-response of a single-source VLC channel, $h(t)$, is composed of two parts as shown in Fig.~\ref{VLC-Response}, i.e., $h(t) = h_{\textmd{LOS}}(t) + h_{\textmd{NLOS}}(t)$. The first term, $h_{\textmd{LOS}}(t)$, is a short pulse with a large peak power received from the LOS path.  This short pulse is followed by a broad pulse with low peak power, $h_{\textmd{NLOS}}(t)$, that corresponds to the multipath effect. The width of these two contributions and the time delay between them, $\tau$, is determined by the room geometry. The second term of the impulse-response causes inter-symbol interference (ISI) in VLC systems. In this work we assume that the duration of the LOS part is short compared with the pulse-width, i.e. $h_{\textmd{LOS}}(t) \sim \delta(t)$, and we approximate $h_{\textmd{NLOS}}(t)$ as a Gaussian function with width $\sigma$. This approximation is accurate for indoor environments with many reflecting surfaces.  LOS reception may not be possible in some parts of the indoor space; the LOS is then blocked, and the impulse-response of the channel becomes $h(t) = h_{\textmd{NLOS}}(t)$.

In a multi-source VLC system, we assume that the data transmitted from all LED arrays in a room are the same and synchronous. For each receiver, the closest LED array is considered the main data source, and the signal received from the direct path of that source is assumed to be the main signal. The signals from other sources that are far from the receiver and the signals reflected from the walls contribute to $h_{\textmd{NLOS}}(t)$.

\subsection{Modulation for VLC}\label{sec:mod}

In order to achieve a data-rate higher than 100 Mbits/s with these LEDs, appropriate coding and modulation must be used. Modified forms of OFDM have been used to achieve a high-speed transmission \cite{OWC-Dimming-OFDM12}. In \cite{OWC-ICTON12} discrete multitone (DMT) OFDM is used to modulate a trichromatic LED with a bandwidth of 180 MHz; for a illumination of 320 lx, data-rates of 397 Mbit/s, 132 Mbit/s and 171 Mbit/s are  reported for the red, green and blue channels, respectively.

PPM is another alternative to modulate white LEDs in indoor wireless systems \cite{OWC-Coded-PPM12}, \cite{OWC-OPPM-PWM12}.
In this work we use expurgated PPM (EPPM) to modulate trichromatic LEDs. EPPM is introduced in \cite{EPPM12} as an alternative technique to PPM to improve the performance of peak-power limited communication systems. We propose to modulate each of the spectral components of a trichromatic LED independently, and at the receiver side we use three photo-detectors to detect the three modulated signals. In EPPM, symmetric balanced incomplete block designs (BIBD) are used as modulated symbols in order to increase the Hamming distance between symbols. Because of the cyclic structure of the BIBD codes, the transmitter and receiver have low complexity and can be implemented using shift registers. A BIBD code is identified by parameters ($Q$,$K$,$\lambda$), where $Q$ is the length, $K$ is the Hamming weight and $\lambda$ is the cross-correlation between two codewords. The $m$th codeword is denoted by the vector $\mathbf{c}_m=(c_{m1}, c_{m2},\dots, c_{mQ} )$, $m = 1, 2, \dots, Q $, where $c_{mi} \in \{0,1\}$. The following relation holds between $\mathbf{c}_{m}$'s \cite{Noshad11}
    \begin{align}\label{Cross Correlation Property}
        \sum_{i=1}^{Q} c_{mi} c_{ni} = \left\{ \begin{array}{lll}
                                        K            &; \, m = n, \\
                                        \lambda      &; \, m \neq n
                                        \end{array}
              \right.  .
    \end{align}
Let $\mathbf{s}_{k}=(s_{k1}, s_{k2}, \dots, s_{kQ}) \in \{\textit{\textbf{c}}_1, \textit{\textbf{c}}_2, \dots, \textit{\textbf{c}}_Q \}$ be the symbol sent in the  $k$th symbol-time.  The signal transmitted by the LED has intensity proportional to $\sum\limits_{k}\sum\limits_{j=1}^Q s_{kj}p(t-kT_s-j\frac{T_s}{Q})$, where $T_s$ is the symbol time and $p(t)$ is the transmit pulse shape.

    \begin{table}[!t]
    \caption{Different BIBD Codes with Various Length to Weight Ratios}\label{BIBD List}
        \vspace*{-0.1 in}
        \begin{center}{\small
            \begin{tabular}{|cc|cc|}\hline
            ($Q$,$K$,$\lambda$) & $Q/K$ &  ($Q$,$K$,$\lambda$) & $Q/K$ \\
            \hline
            (35,17,8)       & 2.058    &            (21,5,1)	    & 4.2    \\
            (11,5,2)        & 2.2      &            (31,6,1)        & 5.167    \\
            (7,3,1)         & 2.33     &            (57,8,1)        & 7.125    \\
            (40,13,4)       & 3.077    &            (91,10,1)       & 9.1    \\
            (13,4,1)        & 3.25     &            (183,14,1)      & 13.07    \\
            (109,28,7)      & 3.89     &            (381,20,1)      & 19.05    \\
            \hline
            \end{tabular}}
        \end{center}
        \vspace*{-0.15 in}
    \end{table}

One of the advantages of EPPM for application in indoor optical wireless systems is the capability of controlling its PAPR. The PAPR of a ($Q$,$K$,$\lambda$)-EPPM is $Q/K$, and this ratio can be controlled by choosing an appropriate BIBD code. There are a vast number of BIBD families with various code-weight to code-length ratios, and they provide a broad range of dimming levels. Table~\ref{BIBD List} lists $Q/K$ ratios for some known BIBD codes \cite{Code-Designs} with PAPR larger than 2.  The complement of a ($Q$,$K$,$\lambda$)-BIBD code is also a BIBD code, with PAPR of $Q/(Q-K)$, and thus these can be used to achieve a PAPR smaller than 2.  For  instance, using just two codes four dimming levels can be obtained.


In \cite{Multilevel-EPPM12}, a multilevel form of EPPM is introduced, and four different modulation schemes are proposed that have a significantly larger spectral efficiency than EPPM due to their large symbol cardinality. Two of these schemes, called type-I multilevel EPPM (MEPPM) and type-II MEPPM, use sums of BIBD codewords to build symbols, and also have a PAPR of $Q/K$. Therefore, similar to EPPM, the PAPR can be controlled in these schemes by using different BIBD codes. Throughout this paper we assume EPPM, but MEPPM could equivalently be used.

\vspace*{-0.1 in}
\subsection{Detection and Demodulation}\label{sec:demod}
    \begin{figure} [!t]
    \begin{center}
    {\includegraphics[width=3.4in]{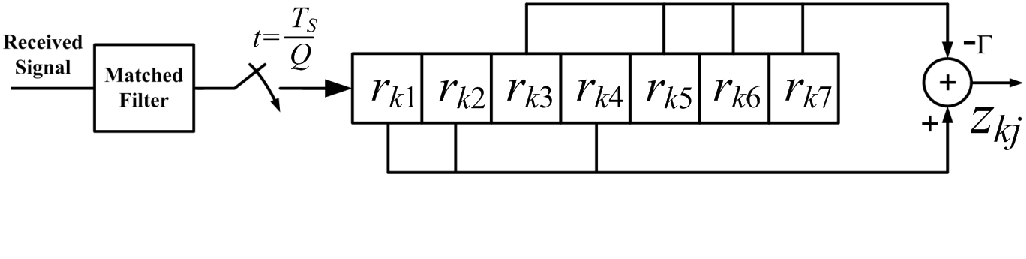}}
    \end{center}
    \vspace*{-0.1 in}
    \caption{Receiver for EPPM using the (7,3,1)-BIBD code.}
    \label{EPPM Decoder}
    \vspace*{-0.1 in}
    \end{figure}
At the receiver, one photodetector is used for each wavelength. An integrate and dump filter is used to sample the received signal after photodetection.  Let $\mathbf{r}_k=(r_{k1},r_{k2},\dots,r_{kQ})$ be the received photon-count in the $Q$ time-slots of symbol $k$. We assume that given a sampling period $T_s/Q$, the VLC channel has an equivalent discrete impulse-response of $h_{\ell}$, for $\forall  \ell \in \mathbb{Z} $, including the effects of the bandwidth-limited transmit pulse-shape and the diffuse multipath channel for a given geometry, calculated from $p(t) \ast h(t)$.

The structure of a simple demodulator for EPPM using a shift-register is shown in Fig.~\ref{EPPM Decoder}. In this receiver, the sampled data in symbol-time $k$, $\mathbf{r}_k$, is stored in a shift register, and then is circulated inside it to generate vector $\mathbf{z}_k=(z_{k1},z_{k2},\dots,z_{kQ})$, $z_{kj}=\langle  \mathbf{r}_k ,\mathbf{c}_j \rangle$, at the output of the differential circuit, where $\langle \mathbf{x}, \mathbf{y} \rangle$ denotes the dot product of the vectors $\mathbf{x}$ and $\mathbf{y}$.
In this figure $\Gamma=\lambda/(K-\lambda)$. The wires of the lower branch are matched to the first codeword of the BIBD code, $\mathbf{c}_1$, and those of the upper branches are matched to its complement. This receiver is equivalent to the correlation decoder, which is the optimum decoder for an additive-white-Gaussian-noise (AWGN) channel in the absence of ISI \cite{EPPM12}. For a system with peak power $P_0$, the output of the decoder looks like PPM but with peak power $KP_0$. The same performance analysis as for PPM can be used for EPPM.

In indoor VLC systems, due to the short distance between transmitter and receiver, the received power is high, and hence, the receiver is shot noise limited.  We show that for ideal non-dispersive channels the receiver in Fig.~\ref{EPPM Decoder} is optimum for shot noise limited Poisson-distributed systems with constant background light.
Letting $h_0=1$ and $h_\ell=0$ for $\ell \neq 0$, the mean of the received photon-count in symbol-time $k$ can be written as $E\{\mathbf{r}_k \} = \Lambda_0 \mathbf{s}_k + \Lambda_b$, where  $\Lambda_0=\eta \frac{P_0}{h \upsilon} \frac{T_s}{Q}$, $P_0$ is the peak received power, $h$ is Planck's constant, $\upsilon$ is the optical frequency, $\eta$ is the efficiency of the photodetector, and $\Lambda_b$ is the photon count caused by background light.
Then the symbol decision in symbol-time $k$, $\hat{m}_k$, using maximum likelihood (ML) detection is
    \begin{align}\label{ML Decoder for EPPM}
        \hat{m}_k &= \arg \max_{1 \le m \le Q} \text{Pr}(\mathbf{r}_k|\mathbf{s}_k=\mathbf{c}_m) \nonumber\\
                &= \arg \max_{1 \le m \le Q} \Bigg( \sum_{j=1}^Q {r_{kj}} \log (c_{mj}\Lambda_0 + \Lambda_b) \nonumber\\
                &\hspace{3.5 cm}- \sum_{j=1}^Q (c_{mj}\Lambda_0+\Lambda_b)  \Bigg).
    \end{align}
Because of the fixed weight of EPPM symbols, $\sum\limits_{j=1}^Q c_{mj}=K$, and, since $c_{mj} \in \{0,1 \}$, the ML decoder reduces to
    \begin{align}\label{ML Decoder for EPPM_short}
        \hat{m}_k &= \arg \max_{1 \le m \le Q} \sum_{j=1}^Q {r_{kj}} c_{mj}.
    \end{align}
As a result, the correlation receiver is the optimum decoder for EPPM in shot-noise limited Poisson-distributed systems.

\begin{figure} [!t]
    \begin{center}
    {\includegraphics[width=3.3in]{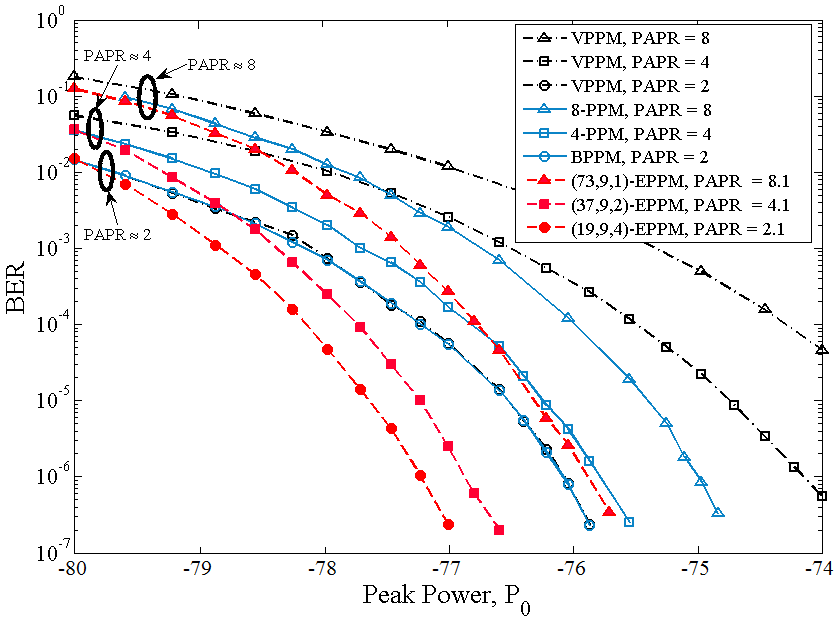}}
    \end{center}
    \vspace*{-0.1 in}
    \caption{Simulated BER versus the received peak power, $P_0$, for EPPM, PPM, and VPPM for different PAPRs.}
    \label{Different PAPR}
    \vspace*{-0.0 in}
    \end{figure}

The simulated bit error rates (BER) of EPPM, PPM and VPPM are compared in Fig.~\ref{Different PAPR} for a PAPR of 2, 4 and 8 at a bit-rate of $200$ Mbits/s for a single color assuming a rectangular transmitted pulse-shape (ideal LED) and an ideal VLC channel without ISI. The shot noise is assumed to be the dominant noise, and the background light is set to $0.1$ $\mu$W. According to these results, for peak-power limited systems, the performance of EPPM surpasses that of PPM and VPPM for all three PAPR levels, since it transmits more than one pulse in each symbol period.

\section{Approaches for Band-limited Sources and Channels}\label{sec:approaches}
In this section, we propose techniques to provide high-speed communications for systems with bandwidth limited sources and channels.
In mildly dispersive channels, the main interference on a symbol is caused by its own pulses. Therefore, and for mathematical simplicity, we ignore the interference between adjacent symbol-times. Also for mathematical convenience, we use cyclic shifts of transmitted vectors instead of their right shifts in our analysis. With these assumptions, given symbol $\mathbf{s}_k$ is sent, and ignoring background light, the mean of the received photoelectron-count at symbol-time $k$ is approximated by the vector $\Lambda_0 \sum\limits_{\ell} h_{\ell} \mathbf{s}{_k^{(\ell)}}$, where the notation $\mathbf{x}{^{(\ell)}}$ denotes the $\ell$th right cyclic-shift of vector $\mathbf{x}$.

\subsection{Interleaved EPPM for Dispersive Channels}\label{sec:interleaving}
As mentioned in Section~\ref{sec:system}, a bit-rate limiting factor in indoor VLC systems is the multipath effect. The multipath effect broadens the transmitted pulses and imposes ISI on the transmitted data. Interleaving is introduced here as a technique to improve the performance of EPPM in multipath channels, which makes EPPM an appealing alternative to PPM for dispersive communication systems.

    \begin{figure} [!t]
    \begin{center}
    {\includegraphics[width=3.3in]{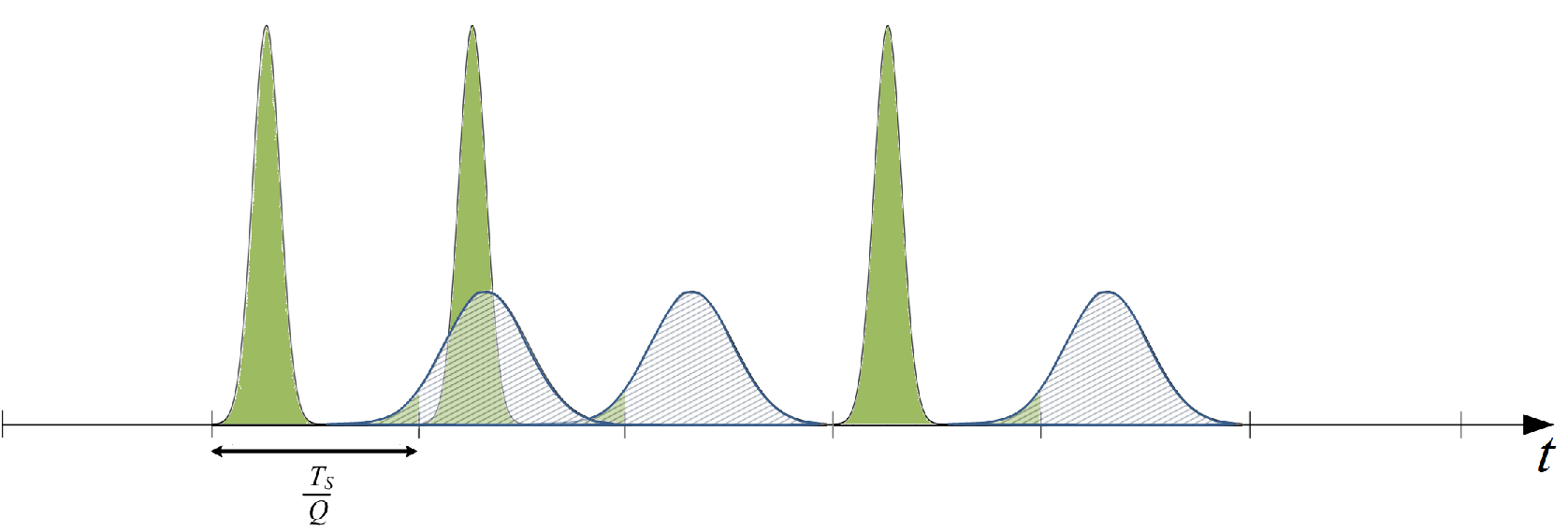}}
    \end{center}
    \vspace*{-0.1 in}
    \caption{Received signal in a dispersive VLC indoor channel. Solid colored areas are the main received signals, and shaded areas are interference signals.}
    \label{Broadened Pulses}
    \vspace*{-0.1 in}
    \end{figure}

In a dispersive channel each pulse is broadened over its neighboring time-slots, and causes interference, as shown in Fig.~\ref{Broadened Pulses}. Therefore, given codeword $\mathbf{c}_m$ is sent, an average signal photoelectron count $\Lambda_0 \sum\limits_{\ell} h_{\ell} \mathbf{c}{_m^{(\ell)}}$ is received at the output of the channel. Since the symbols of EPPM are cyclic shifts, $\mathbf{c}{_m^{(\ell)}} = \mathbf{c}{_{m+\ell}}$, and hence, the received signal can be represented as $ \sum\limits_{\ell} h_{\ell} \mathbf{c}{_{m+\ell}}$. This means that the expected $j$th output of the decoder, $z_{kj}$, in Fig.~\ref{EPPM Decoder} ignoring background light is \cite{EPPM12}
    \begin{align}\label{}
        E\big\{z_{kj}|&\mathbf{s}_k=\mathbf{c}_m \big\} \nonumber\\
        &= \Lambda_0 \bigg( \Big\langle \mathbf{c}_j, \sum_{\ell} h_{\ell} \mathbf{c}{_{m+\ell}}\Big\rangle - \Gamma \Big\langle\overline{\mathbf{c}_j}, \sum_{\ell} h_{\ell} \mathbf{c}{_{m+\ell}}\Big\rangle \bigg) \nonumber\\
        &= \Lambda_0 K h_{m-j}.
    \end{align}
Thus the distance between adjacent symbols decreases at the output of the channel and this increases the error probability.

Since the EPPM symbols are cyclic shifts, the delay between LOS and NLOS responses, $\tau$, only affects which of the symbols the interference falls upon, and has no effect on the performance as long as $\tau>\frac{T_s}{Q}$. Therefore, the magnitude of $h_{\textmd{NLOS}}$ and its width, $\sigma$, are the only parameters that influence the amount of interference endured.

    \begin{figure} [!t]
    \begin{center}
    {\includegraphics[width=3.2in]{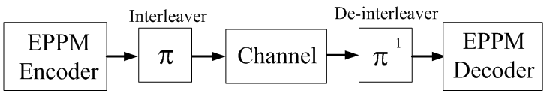}}
    \end{center}
    \vspace*{-0.1 in}
    \caption{Schematic view of an EPPM system using an interleaver and a de-interleaver for dispersive channels. }
    \label{EPPM Interleaver}
    \vspace*{-0.1 in}
    \end{figure}
A symbol-length interleaver and deinterleaver can be used in the transmitter and receiver to diminish the impact of ISI, as shown in Fig.~\ref{EPPM Interleaver}. This interleaver can be considered as a permutation matrix, $\pi$, with element $(i,j)$ denoted as $\pi_{ij}$, and the deinterleaver would be its inverse, $\mathbf{\pi}^{-1}$. So $\mathbf{c}_m \pi$ is transmitted instead of $\mathbf{c}_m$, i.e., the transmitted pulses in the EPPM symbol are rearranged. The code-length and code-weight of a ($Q$,$K$,$\lambda$)-EPPM after the interleaver remain as $Q$ and $K$, respectively, and hence, the interleaver does not change the PAPR. At the receiver side, the inverse of the permutation matrix is applied on the received vector and then the decoder in Fig.~\ref{EPPM Decoder} is used to decode the received signal. Thus, in a noiseless non-dispersive channel, the output of the channel is $\mathbf{c}_m \pi$, and $\mathbf{c}_m \pi \pi^{-1} = \mathbf{c}_m$ is sent to the EPPM decoder.
    \begin{figure} [!t]
    \begin{center}
    {\includegraphics[width=3.4in]{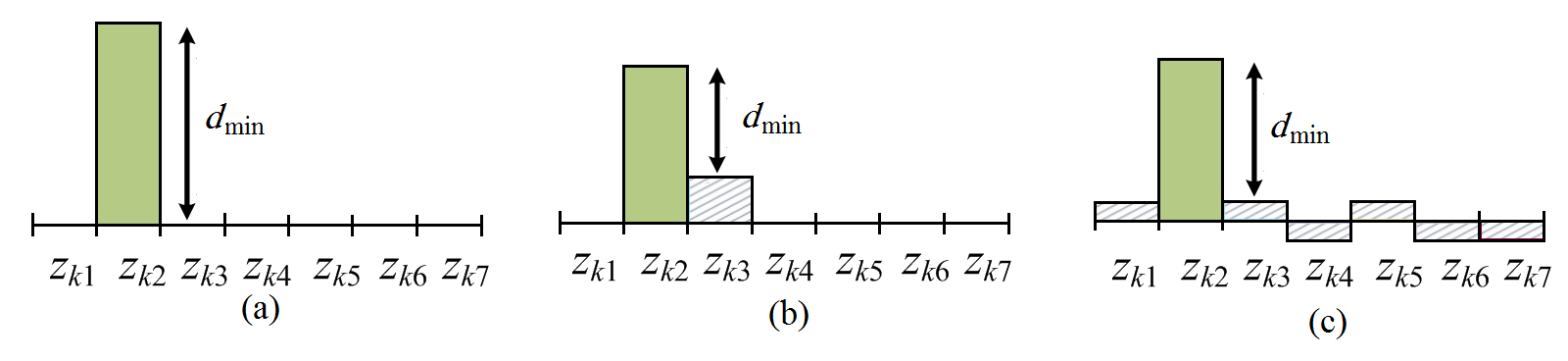}}
    \end{center}
    \vspace*{-0.2 in}
    \caption{The decoded signal and the minimum distance between symbols, $d_{\min}$, for (a) ideal channel and dispersive channels (b) without interleaver, and (c) with interleaver.}
    \label{Interleaver Effect}
    \vspace*{-0.1 in}
    \end{figure}

Now let us consider the performance of this technique in dispersive channels. Still assuming $\mathbf{c}_m$ is transmitted, in this case the noiseless output of the channel is proportional to $\sum\limits_{\ell} h_{\ell} \left(\mathbf{c}_m \pi \right)^{(\ell)}$, and $h_0 \mathbf{c}_m + \sum\limits_{\ell \neq 0} h_{\ell} \left(\mathbf{c}_m \pi \right)^{(\ell)} \pi^{-1}$ is sent to the EPPM decoder. Therefore, the expected value of the $j$th output of the decoder is
    \begin{align}\label{Mean of z_j when interleaver is used}
        E\big\{&z_{kj}|\mathbf{s}_k=\mathbf{c}_m  \big\} =  \Lambda_0 \bigg( h_0 K \delta_{jm}+ \nonumber\\
        & (1+\Gamma)\sum_{\ell \neq 0} h_{\ell} \left(\left\langle \mathbf{c}_j, \left(\mathbf{c}_m \pi \right)^{(\ell)} \pi^{-1} \right\rangle -K \right) \bigg),
    \end{align}
where
    \begin{align}\label{}
        \delta_{jm} = \left\{ \begin{array}{lll}
                                        1            &; \, j = m, \\
                                        0      &; \, j \neq m
                                        \end{array}
              \right.  .
    \end{align}
%
%
Fig.~\ref{Interleaver Effect} shows the output of the decoder for both ideal (Fig.~\ref{Interleaver Effect}-(a)) and dispersive (Fig.~\ref{Interleaver Effect}-(b)) channels, and the effect of the interleaver on the decoded signal. In this example, the tails of the received pulses due to the channel interfere with one adjacent symbol, and, therefore, the minimum distance between decoded signals is considerably reduced compared with the dispersion-less case. Adding an interleaver spreads the interference over all symbols, as shown in Fig.~\ref{Interleaver Effect}-(c), and hence, the minimum distance between decoded signals is increased.

The same interleaver/deinterleaver technique can also be used for multilevel-EPPM to increase the minimum distance between MEPPM symbols in dispersive channels. In \cite{Multilevel-EPPM12}, MEPPM is shown to achieve higher spectral-efficiencies compared to EPPM due to its larger constellation size, and thus, longer symbol-time. Hence, interleaved MEPPM can achieve an even-lower error probability in bandwidth-limited channels.

According to Section~\ref{sec:demod}, the demodulator in Fig.~\ref{EPPM Decoder} computes $\hat{m}_k=\arg \max\limits_{1 \leq m \leq Q} z_{km}$. Defining $d_{mj}:=E\{(z_{km}-z_{kj})|\mathbf{s}_k = \mathbf{c}_m\}$, we assume that the error probability between symbols $m$ and $j$ can be expressed as $f(d_{mj})$, where $f(.)$ is a monotonically decreasing function. Then the symbol error probability can be approximated by \cite{EPPM12}
\begin{equation}\label{eq:Ps}
P_s\approx \frac{1}{Q} \sum_{m=1}^Q \sum_{\substack{j=1\\j \neq m}}^Q f(d_{mj}).
\end{equation}
Therefore, in high signal to noise (SNR) regimes, we can write $P_s \simeq M_{d_{\min}} f(d_{\min})/Q$, where $d_{\min} := \min\limits_{ m\neq j}{d_{mj}}$ and $M_{d_{\min}}$ is the number of $(m,j)$ pairs with $d_{mj}=d_{\min}$.

The optimal permutation matrix that minimizes the symbol error probability is the one that not only maximizes $d_{\min}$, but also minimizes $M_{d_{\min}}$. We first calculate a lower bound on $P_s$ of interleaved EPPM, and then use binary linear programming (BLP) to find the best arrangement for the interfering pulses for a given channel. Note that the optimum permutation matrix for an EPPM scheme with a given BIBD code may not be optimum for multilevel-EPPM that uses the same BIBD code.\\

\subsubsection{Lower Bound on $P_s$ of Interleaved EPPM}\label{sec:bound}\mbox{}\\

Letting $A_{j,m,\ell} = \left\langle \mathbf{c}_j,\left( \left(\mathbf{c}_m \pi \right)^{(\ell)} \pi^{-1}\right) \right\rangle$,
    \begin{align}\label{minimum distance}
        d_{mj}  \propto h_0 K -  (1+\Gamma) \sum_{\ell \neq 0} h_{\ell} \left( A_{j,m,\ell} -A_{m,m,\ell} \right), \, m \neq j.
    \end{align}
Thus maximizing $d_{\min}$ is equivalent to minimizing $\max\limits_{m,j} \sum\limits_{\ell \neq 0} h_{\ell}( A_{j,m,\ell} -A_{m,m,\ell} )$. To obtain a lower bound on $P_s$, we minimize $\max_{m,j} ( A_{j,m,\ell} -A_{m,m,\ell} )$ for each $\ell \neq 0$. To this end, we rewrite $A_{j,m,\ell}$ as $A_{j,m,\ell} = (\mathbf{c}_m \pi ) \mathbf{I}_{\ell} (\mathbf{c}_j \pi)^{\text{T}}  $, where $\mathbf{I}_{\ell}$ is the $\ell$th right cyclic shift of the identity matrix. We can then define the $Q\times Q$ matrix $\mathbf{A}_{\ell}$ with $(j,m)$ element equal to $A_{j,m,\ell}$, which can be represented as
    \begin{align}\label{}
        \mathbf{A}_{\ell} = \mathbf{C} \, \pi \, \mathbf{I}_{\ell} \, \pi^{\text{T}} \, \mathbf{C}^{\text{T}},
    \end{align}
where $\mathbf{C}$ is the code matrix with $m$th row equal to $\mathbf{c}_m$, for $m=1,2,\dots,Q$. We first prove the following lemma which is then used to derive constraints on the elements of $\mathbf{A}_{\ell}$.\\

\noindent \textbf{Lemma}: Let $\pi$ be a permutation matrix, and define $\mathbf{P}_{\ell}:=\pi \, \mathbf{I}_{\ell} \, \pi^{\text{T}}$. Then the diagonal elements of $\mathbf{P}_{\ell}$ are all zeros for $\ell \neq 0$.

\noindent \textbf{Proof}:  $\mathbf{P}_{\ell}$ is clearly also a permutation matrix. Let $p_i$ and $\pi_i$, $i=1,2,\dots,Q$, denote the positions of the `1' in the $i$th row of $\mathbf{P}_{\ell}$ and $\pi$, respectively. Therefore, $\mathbf{P}_{\ell}=\pi \, \mathbf{I}_{\ell} \, \pi^{\text{T}}$ implies that $\pi_{p_i}=\pi_i+1$, and therefore  $p_i\neq i$ (the diagonal elements of $\mathbf{P}_{\ell}$ are zero). $\blacksquare$

In order to calculate a lower bound on $P_s$, we prove the following theorem.\\

\noindent\textbf{Theorem}: The following three constraints hold for $A_{j,m,\ell}$'s:

(i) $\sum\limits_{j=1}^Q A_{j,m,\ell}=K^2$ and $\sum\limits_{m=1}^Q A_{j,m,\ell} = K^2$,

(ii) $\sum\limits_{m=1}^Q A_{m,m,\ell} =\lambda Q$.

\noindent\textbf{Proof}: For (i), using (\ref{Cross Correlation Property}) we obtain
    \begin{align}\label{constraint on sum of columns}
        \sum_{j=1}^Q A_{j,m,\ell} = \bigg\langle \sum_{j=1}^Q \mathbf{c}_j,\left( \left(\mathbf{c}_m \pi \right)^{(\ell)} \pi^{-1}\right) \bigg\rangle = K^2,
    \end{align}
and similarly for the summation over the second index.
To prove (ii), the sum of diagonal elements in $\mathbf{A}_{\ell}$ can be written as
    \begin{align}\label{}
        \sum_{m=1}^Q A_{m,m,\ell} &=\sum_{m=1}^Q \mathbf{c}_m \mathbf{P}_{\ell} \, \mathbf{c}{_m^{\text{T}}}\nonumber\\
        &=  \mathbf{c}_1 \left(\sum_{m=1}^Q \mathbf{I}_{m}\, \mathbf{P}_{\ell} \, \mathbf{I}{_{m}^{\text{T}}} \right) \mathbf{c}{_1^{\text{T}}},
    \end{align}
since $\mathbf{c}_m = \mathbf{c}_1 \mathbf{I}_{m}$. As $\mathbf{P}_{\ell}$ is a permutation matrix, $\mathbf{T} := \left(\sum\limits_{m=1}^Q \mathbf{I}_{m}\, \mathbf{P}_{\ell} \, \mathbf{I}{_{m}^{\text{T}}} \right)$ is a circulant matrix with first column labeled $(T_1, T_2, \dots, T_Q)^{\text{T}}$, and $T_1+T_2+\dots+T_Q=Q$. From (\ref{Cross Correlation Property}), $\sum\limits_{m=1}^Q A_{m,m,\ell}=KT_1+\lambda(T_2+T_3+\dots+T_Q)$, and hence,
    \begin{align}\label{}
        \sum_{m=1}^Q A_{m,m,\ell} &=\lambda Q + (K-\lambda)T_1.
    \end{align}
From the Lemma, $T_1=0$, and hence (ii) holds. $\blacksquare$

From the Theorem, the ideal interleaver is the one that generates $\mathbf{A}_{\ell}$'s, $\ell=1,2,\dots,Q-1$, that have $(K-\lambda)$ nondiagonal elements with value $(\lambda+1)$ in each row and each column, and the rest of the elements are $\lambda$. Hence, a lower bound on $P_s$ is given by
    \begin{align}\label{constraint on diagonal}
        P_s \geq (K-\lambda)f\bigg( \Lambda_0 K \big[h_0  - \frac{1}{K-\lambda} \sum_{\ell \neq 0} h_{\ell}\big] \bigg).
    \end{align}
    \begin{figure} [!t]
    \vspace*{-0.0 in}
    \begin{center}
    {\includegraphics[width=3.2in]{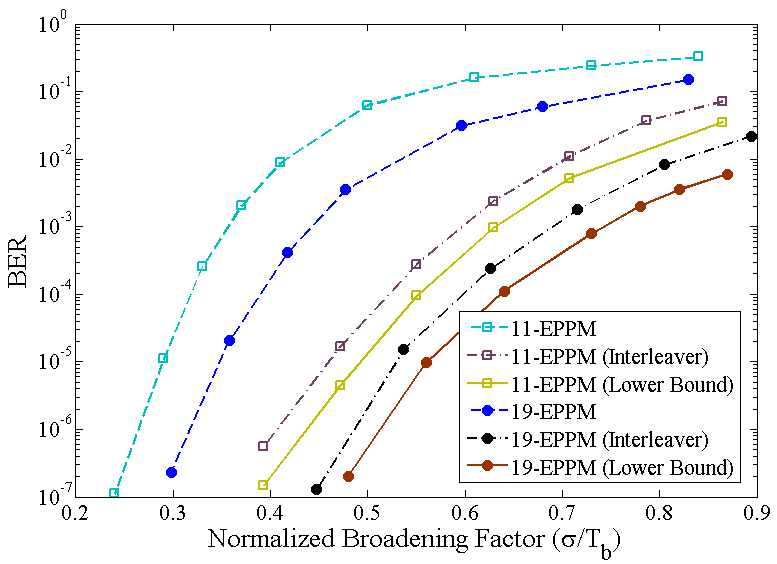}}
    \end{center}
    \vspace*{-0.1 in}
    \caption{Simulated BER vs. normalized broadening factor ($\sigma/T_b$) for EPPM with and without interleaving using (11,5,2) and (19,9,4) BIBD codes in a VLC channel with LOS and NLOS paths.}
    \label{BER Interleaved-EPPM-LOS}
    \end{figure}

\subsubsection{Numerical Results}\label{sec:interleaved_results}\mbox{}\\

Although the results above apply to any dispersive channel, in this section we provide numerical results specifically for a VLC system. In all results in this section we make the following assumptions: the VLC system is affected by multipath, shot noise and background noise; the bit rate is fixed to be $R_b=1/T_b=200$ Mbits/s per color; a photo-detector with an effective area of 0.1 cm$^2$ and quantum efficiency of 0.7 is used for photo-detection; the best permutation matrix for the interleaver for each code is found using BLP as shown in the appendix; simulations use perfect symbol synchronization; no equalizer is used at the receiver; the VLC system receiver is exposed to 0.1 $\mu$W of background light.

Performance results using an (11,5,2)-EPPM and a (19,9,4)-EPPM with and without interleaving are plotted versus the normalized broadening factor, $\sigma/T_b$, in Fig.~\ref{BER Interleaved-EPPM-LOS}.
We use the results of \cite{OWC-Channel05}, \cite{OWC-Channel11} and \cite{OWC-Channel93} to simulate $h(t)$, the impulse-response shown in Fig.~\ref{VLC-Response}, where the channel is a combination of LOS and NLOS paths components. The delay between LOS and NLOS responses, $\tau$, is assumed to be small compared to the symbol-time $T_s$ but large compared with the slot time $T_s/Q$.  The energy per bit received via the LOS path is assumed to be $5\times10^{-16}$ J, which corresponds to an SNR of 15 dB.
The NLOS impulse response is approximated by a Gaussian pulse with a fixed energy, and the received energy from the NLOS path is assumed to be $5\times10^{-16}$ J.
According to these results, the energy of the interference signal is divided between a larger number of time-chips for a longer code-length, and it therefore achieves a lower BER compared to the shorter code-length. Moreover, for both codes the BER for EPPM is reduced using an interleaver with a good permutation matrix. Lower bounds on the BER of the interleaved EPPM show the ultimate BERs that can be achieved using interleaving.
%
    \begin{figure} [!t]
    \vspace*{-0.0 in}
    \begin{center}
    {\includegraphics[width=3.2in]{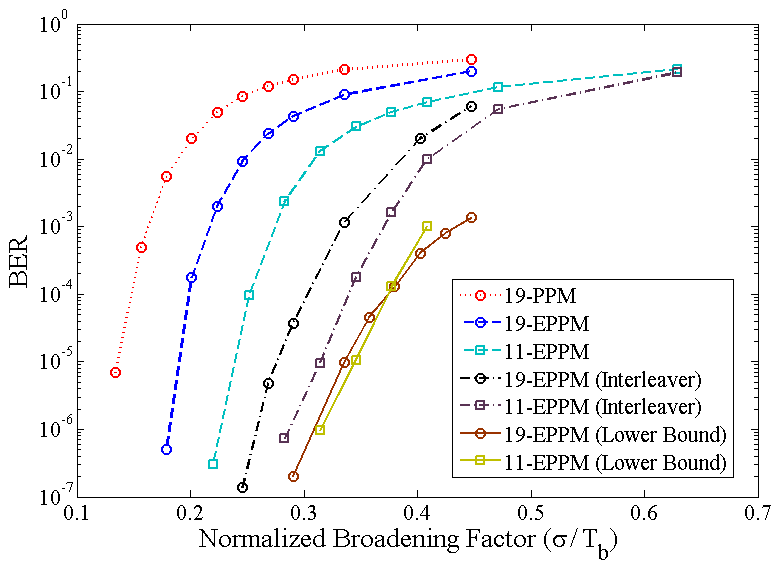}}
    \end{center}
    \vspace*{-0.1 in}
    \caption{Simulated BER vs. normalized broadening factor for PPM and EPPM with and without interleaving using (11,5,2) and (19,9,4) BIBD codes in a VLC channel with blocked LOS path (only the NLOS portion of the impulse response).}
    \label{BER Interleaved-EPPM-NLOS}
    \end{figure}
    \begin{figure} [!t]
    \vspace*{-0.0 in}
    \begin{center}
    {\includegraphics[width=3.2in]{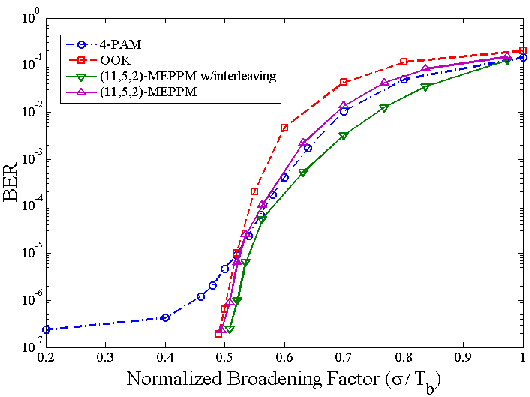}}
    \end{center}
    \vspace*{-0.1 in}
    \caption{Simulated BER vs. normalized broadening factor ($\sigma/T_b$) for OOK, 4-ary PAM and 11-level type-II MEPPM with and without interleaving.}
    \label{BER Interleaved-MEPPM}
    \end{figure}

Fig.~\ref{BER Interleaved-EPPM-NLOS} shows the interleaving effect on the BER of EPPM for a receiver that does not have a LOS path to the LEDs. In this figure the BER is again plotted versus the normalized broadening factor, where the channel is now assumed to have a Gaussian-shaped impulse-response. Here the energy received from the NLOS is assumed to be a constant $5\times10^{-16}$ J as $\sigma$ increases. According to these results, the BER for EPPM is lower than for PPM with the same length. The lower bound is tighter because the interference is less severe than in Fig.~\ref{BER Interleaved-EPPM-LOS}.

In Fig.~\ref{BER Interleaved-MEPPM}, the BER of 11-level type-II MEPPM using a (11,5,2)-BIBD code is plotted versus the normalized broadening factor and  compared to on-off-keying (OOK) and 4-ary pulse amplitude modulation (PAM), assuming the same channel as for Fig.~\ref{BER Interleaved-EPPM-LOS}. Although 4-PAM has a better performance for high dispersive channels compared to OOK and MEPPM without interleaving, its BER is large for channels with weaker multipath effect. The BER of MEPPM is improved using interleaving, and its performance surpasses 4-PAM in all cases tried.

\subsection{Overlapped EPPM for Bandwidth-Limited Sources}\label{sec:overlapping}
As mentioned in Section~\ref{sec:system}, white LEDs used in VLC systems have a long relaxation time, and therefore limit the transmission rate. Overlapped PPM (OPPM) was first proposed for general optical communication systems with bandwidth-limited sources in \cite{OPPM-84} to increase the data-rate by spreading the pulses over multiple time-chips. OPPM has recently been proposed specifically for indoor VLC application \cite{OWC-OPPM-PWM12}, since it can increase the transmission rate of white LEDs. Using the same idea, we combine the overlapped pulses technique with EPPM in order to increase the symbol-rate in VLC systems. We call the new scheme overlapped EPPM (OEPPM).
    \begin{figure} [!t]
    \vspace*{-0.0 in}
    \begin{center}
    \scalebox{0.18}{\includegraphics{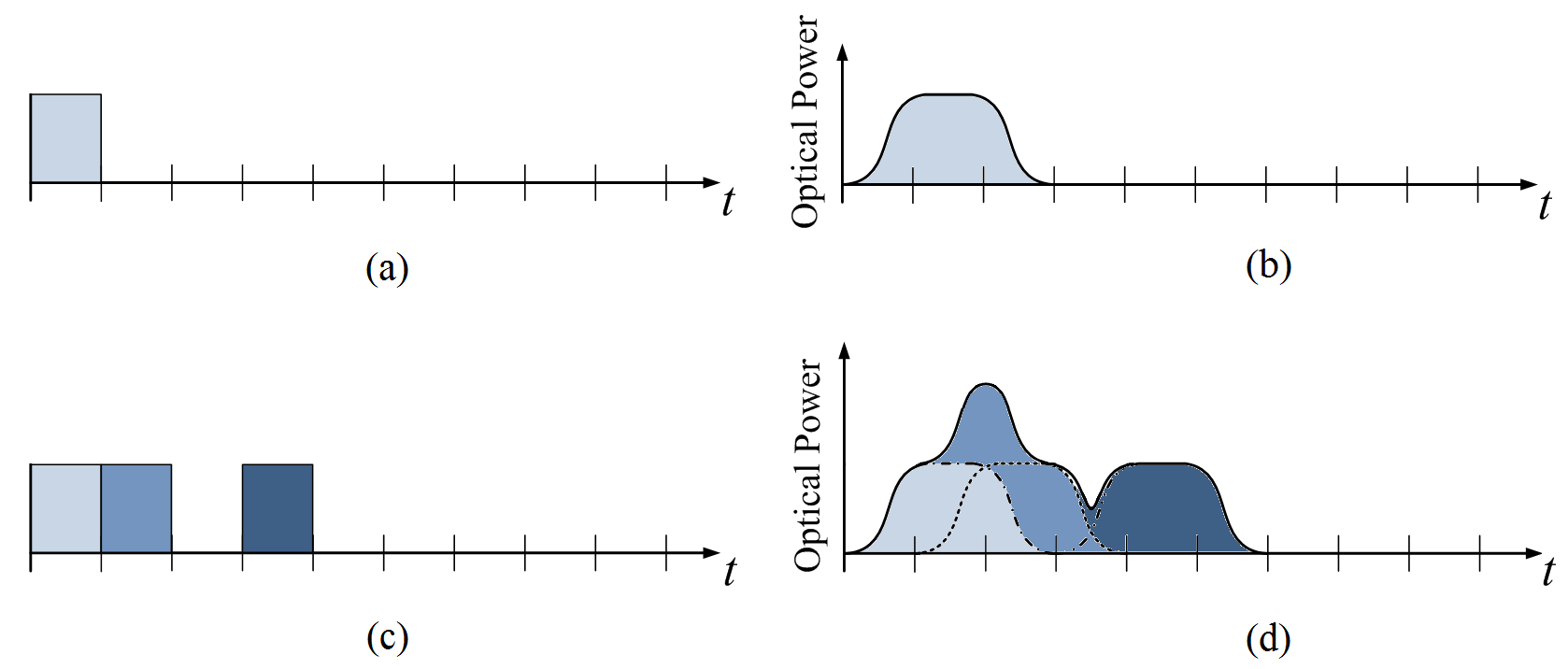}}
    \end{center}
    \vspace*{-0.1 in}
    \caption{(a) A rectangular pulse input, and (b) the corresponding response of an LED with a sixth order Bessel filter model. (c) The first codeword of a (7,3,1)-BIBD code, and (d) its corresponding generated overlapping-EPPM symbol.}
    \label{Overlapping-EPPM}
    \end{figure}

In this scheme, we assume that the pulses generated by the LEDs are $v$ times wider than a time-slot. For an OEPPM that uses a ($Q$,$K$,$\lambda$)-BIBD code to modulate the symbols, each symbol is divided into $(Q+v-1)$ equal time-slots, and $K$ pulses with width $v$ time-slots are transmitted in each symbol period. Fig.~\ref{Overlapping-EPPM} shows the generation of a OEPPM symbol using a (7,3,1)-BIBD codeword and overlapping length of $v=3$. The response of each LED to a single rectangular input signal (Fig.~\ref{Overlapping-EPPM}-(a)) is shown in Fig.~\ref{Overlapping-EPPM}-(b). The first codeword of the (7,3,1)-BIBD code and its corresponding OEPPM symbol are shown in Fig.~\ref{Overlapping-EPPM}-(c) and Fig.~\ref{Overlapping-EPPM}-(d), respectively. The rising edges of the optical pulses are determined by the ``1"s in the corresponding BIBD codeword.

The OEPPM waveform is multilevel, and requires multiple LEDs to generate if each LED is on-off pulse modulated. An LED array with $K$ pulse-modulated LEDs can implement any OEPPM symbol: each LED in the array is modulated by one pulse of the generating BIBD, and stays on for $v$ time-slots corresponding to its limited bandwidth.  The time between two successive ``1"s in a BIBD codeword could be larger than $v$ time-slots, and therefore, the number of levels in its corresponding OEPPM symbol could be smaller than $K$, in fact no larger than $\min(v,K)$. Consequently, OEPPM can always be implemented using no more than $\min(v,K)$ pulse-modulated LEDs. Assuming that at most $N_{\textmd{LED}}$ pulses overlap, requiring $N_{\textmd{LED}}$ LEDs to generate the waveform, the resulting minimum Euclidean distance between transmitted symbols is $\frac{T_{\textmd{LED}} P_0}{v}\frac{\sqrt{K-\lambda}}{N_{\textmd{LED}}}$, where $P_0$ is the peak power of the entire LED array, and $T_{\textmd{LED}}$ is the LED impulse response duration. The PAPR of this modulation is $\frac{N_{\textmd{LED}}(Q+v-1)}{vK}$.

    \begin{figure} [!t]
    \vspace*{-0.0 in}
    \begin{center}
    \scalebox{0.40}{\includegraphics{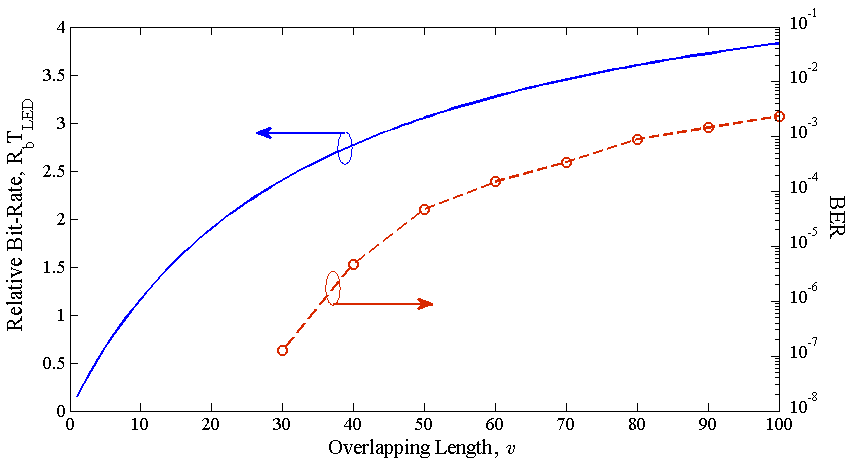}}
    \end{center}
    \vspace*{-0.1 in}
    \caption{The maximum attainable relative bit-rate, $R_b T_{\textmd{LED}}$ (solid line), for $Q=35$, and its corresponding simulated BER (dashed line) versus overlapping length, $v$.}
    \label{Optimum Q for v OEPPM}
    \end{figure}

The bit-rate of a ($Q$,$K$,$\lambda$)-OEPPM is the same as for $Q$-ary OPPM. Given that each color diode of the LED has a impulse-response duration of $T_{\textmd{LED}}$, the bit-rate using OEPPM with $v$ overlapping time-slots, per color, is given by
    \begin{align}\label{OEPPM Rate}
        R_{b,\text{OEPPM}} = \frac{\log_2 Q}{T_{\textmd{LED}}} \frac{v}{Q+v-1}.
    \end{align}
By increasing $v$ the bit-rate can increase, up to a limit of $\frac{\log_2 Q}{T_{\textmd{LED}}}$. Therefore, OEPPM schemes with larger $Q$'s can achieve higher data-rates.
Fig.~\ref{Optimum Q for v OEPPM} shows the maximum achievable relative bit-rate, $R_b T_{\textmd{LED}}$ for $Q=35$ as a function of $v$.  The corresponding simulated shot-noise-limited BER is also shown. In these results, the peak received power is assumed to be 40 $\mu$W, and the LED is modeled as a sixth order Bessel filter with $T_{\textmd{LED}}=20$ ns. For these values a bit-rate of $210$ Mb/s per color is achievable with a BER of $10^{-3}$.

The overlapping-pulse technique can also be utilized with the MEPPM scheme to further increase the data-rate possible with bandwidth-limited LEDs. The bit-rate of type-I and type-II MEPPM schemes utilizing $v$ overlapped pulses are
    \begin{align}\label{}
        R_{b,\text{OMEPPM-I}} = \frac{\log_2 {{Q}\choose{N}}}{T_{\text{LED}}} \frac{v}{Q+v-1},
    \end{align}
and
    \begin{align}\label{}
        R_{b,\text{OMEPPM-II}} = \frac{\log_2 {{Q+N}\choose{N}}}{T_{\text{LED}}} \frac{v}{Q+v-1},
    \end{align}
respectively. Accordingly, for a fixed $Q$ and arbitrarily large $v$, the maximum attainable relative bit-rate, $R_b T_{\textmd{LED}}$, for each scheme is ${\log_2 {{Q}\choose{N}}}$ and ${\log_2 {{Q+N}\choose{N}}}$.

We can combine the pulse-overlapping and symbol interleaving for systems with both limited source and channel bandwidths. The pulse-overlapping can increase the data-rate of the transmitter, while interleaving reduces the channel-imposed interference between pulses.

\section{Conclusion}\label{sec:conclusion}

In this Part I of a two-part paper, several techniques using EPPM modulation are proposed to increase the data-rate in VLC systems. Dimming of the lighting system can be done by changing the BIBD code, achieving an arbitrary level of illumination by using an appropriate code-weight and code-length. The correlation decoder is shown to be optimal for shot-noise limited systems. Interleaving and overlapping techniques are introduced to increase the bit-rate of EPPM in systems with bandwidth-limited channels and sources, respectively. These techniques can also be used with MEPPM to further increase the spectral efficiency.  In Part II of the paper, we develop multiple-access techniques using EPPM for indoor VLC networks.

\section{Acknowledgment}
This research was funded by the National Science Foundation (NSF) under grant number ECCS-0901682.

%
\appendices
\numberwithin{equation}{section}
%

\section*{Appendix \\ Optimal Interleaver Design}

In this appendix we describe how to find the best interleaver for a given channel impulse response.  Let $B_{i,j}:=\sum\limits_{\ell \neq 0} h_{\ell} A_{i,j,\ell}$, for $\forall i,j$, and $b_{\max}=\max\limits_{i,j} B_{i,j}$. According to (\ref{minimum distance}), the optimum $\pi$ is the one that: 1) minimizes $b_{\max}$ (i.e., maximizes $d_{\min}$), and 2) minimizes $\big|\{B_{i,j}=b_{\max}|1\leq i,j\leq Q\}\big|$ (i.e., minimizes $M_{d_{\min}}$). To find the optimum $\pi$, we solve two successive BLPs, with the same constraints but with different objective functions. In the first BLP, the goal is to minimize $b_{\max}$, and then in the second BLP we minimize $\big|\{B_{i,j}=b_{\max}|1\leq i,j\leq Q\}\big|$.
In both programs, the sum of elements in each row and each column is set to 1 as equality constraints. The variables must also satisfy the following quadratic inequality constraint for $\forall m,j$:
    \begin{align*}\label{}
        \mathbf{c}_m \, \pi \, \mathbf{I}' \, \pi^{\text{T}} \, (\mathbf{c}_j-\mathbf{c}_m)^{\text{T}} \leq b_{\max},
    \end{align*}
where $ \mathbf{I}'=\sum\limits_{\ell} h_{\ell} \mathbf{I}_{\ell}$. These binary quadratic constraints can be reduced to binary linear constraints by representing cross products by binary variables as explained in \cite[pp. 67]{Quadratic2Linear}. This introduces new linear constraints involving new and old variables.

\bibliographystyle{IEEEtran}
\bibliography{EPPM}

\end{document}